# An approach to control collaborative processes in PLM systems

Soumaya El Kadiri[1], Philippe Pernelle[2], Miguel Delattre[3], Abdelaziz Bouras[4],

(1, 3, 4) Université Lumière Lyon 2 – Laboratoire LIESP - IUT Lumière – 160 Bd de l'Université, 69676 BRON Cedex – France, Email : [soumaya.el-kadiri] [miguel.delattre] [abdelaziz.bouras] @univ-lyon2.fr
(2) Université Claude Bernard Lyon 1 – Laboratoire LIESP – 69622 Villeurbanne – France, Email : philippe.pernelle@iutb.univ-lyon1.fr

## Abstract

Companies that collaborate within the product development processes need to implement an effective management of their collaborative activities. Despite the implementation of a PLM system, the collaborative activities are not efficient as it might be expected. This paper presents an analysis of the problems related to the collaborative work using a PLM system. From this analysis, we propose an approach for improving collaborative processes within a PLM system, based on monitoring indicators. This approach leads to identify and therefore to mitigate the brakes of the collaborative work.

**Keywords:** PLM, collaborative processes, collaborative indicators, tracks, observation

## 1. Introduction

Product Lifecycle Management systems are used in industrial enterprises wishing to manage their products data and knowledge, in all phases of their lifecycle (design, manufacturing, recycling, etc). Integrating a PLM system is a strategic approach for industrial groups that aim to improve their product development. Through a set of tools, the enterprise ensures the coordination management and traceability for all information and activities related to the product lifecycle. However, SMEs remain reluctant to implement such systems. From an organizational point of view the small structure of the SME leads to combine various responsibilities. Consequently, informal relationships (internal and external) take an important place within organization. From a process point of view, processes in the SME are characterized by a rigid formalization at a macro level, and a very flexible and non-formalization of the detailed processes, in order to allow their informal relationships [1]. In such context, the real effects of collaboration are not perceivable in spite of the SMEs willingness to better collaborate. Furthermore, some reverse effects may take place, inducing therefore break points.

In this paper we aim to propose an approach for improving collaborative activities (internal and external) with a PLM in a small and medium sized company.

This paper is divided into two sections. In the first section we present the context of our research work, particularly the ANCAR-PLM[1] project. The objective of this project is to contribute to the study and the implementation of methodologies integration and deployment of solutions. These solutions, usually called information systems for PLM, allow managing all information flows around the product/service in all phases of its lifecycle and with all the project partners. PLM systems are primarily used, in order to optimize existing products (capitalization, delays reduction, etc.) or to develop new products. In this context we address specifically the issues of collaborative work within PLM systems in SMEs.

---

[1] ANalyse, CARactérisation et mises en oeuvre de solutions de gestion de cycles de vie de produits et de services (PLM) (http://iutcerral.univ-lyon2.fr/AncarPLM/)

In the second section, we present an approach which aims to limit the brakes related to the collaborative work. This approach consists in two main actions: the first action explains the construction of some monitoring indicators leading to analyze and to identify fluidity lacks (dysfunctions) or break points. The second one proposes a meta model extension to support indicators as well as actions on the processes to limit break points previously identified.

## 2. Collaborative work within PLM systems

Collaboration is a process of participation through which people, groups, and organisations work together to achieve desired results. Collaboration can occur among individuals, groups, or organisations at the same time (synchronous) or with a time delay (asynchronous). Collaboration can also occur between people located in the same place or separated by physical distance [2].

Collaboration occurs at various levels [2]:
- *Informal Collaboration* - This is the simplest level of collaboration, involving activities which are unstructured and informal. Examples of this level of collaboration would include one-to-one communication, discussion groups, and one-off meetings.
- *Process / Project Collaboration* - The next level of collaboration comprises processes that are more structured in nature, which have defined start and end points, as well as a defined flow of events between the two.
- *Extended Collaboration* - The third level of collaboration involves activities that extend beyond the enterprise to include customers, partners, and vendors. Activities in this category would include interaction with customer focus groups, product design sessions with vendors, and delivery of services or products by partners.

Most of industrial activities take place within collaboration with multiples partners (subcontractor, customers …). At the same time this situation leads to many brakes that require, among others, the use of regulation mechanisms and tools.

### 2.1. Collaboration in industrial context

One of the goals of PLM to reduce the product life-cycle is to foster collaboration among different actors involved in product development processes. Thanks to the standard tools in the PLM systems (forum, workflow, messaging…), users have the means to support collaborative projects. These standards are the result of the CSCW[2] researches, which propose a methodological framework to facilitate collaborative work among a group of actors. One of the aspects covered by the CSCW is the interactions analysis between actors which led to collaborative activities segmentation. Moreover this work has resulted in the classical distinction between activities: communication, coordination, collaboration.

This segmentation is incomplete in the context of an industrial activity leading to new products development. Indeed, it is advisable to have an organizational approach regarding to the constraints induced on the collaboration modes and tools (notably processes).

---

[2] Computer Supported Cooperative Work

The consideration of the organizational context is an essential element of differentiation. Indeed, according to the industrial structure (international groups, networked enterprise, sub-contracting, design outsourcing, etc) collaborations between actors are subjected to different constraints. We can, for instance, quote:
- The hierarchical constraints that typically impose restrictions if people are not on the same level.
- The functional constraints established by the enterprise or by the external constraints, such as specific certification or rules related to a specific sector (chemistry, food processing industry) that imposes to respect some procedures.
- The communautarist constraints which are more diffuse but always present in all areas and whatever activity, inducing by the way the feeling of membership to a group of actors rather than to another.
- The customer constraints depend on the degree of his involvement in the company's activity. In the case of outsourcing, the customer is very active and collaborative activities are generally determined by the following paradox: satisfying the customer by providing him the maximum of information without delivering the expertise that maintains the company's sustainability.

Some of these constraints affect partially, but in explicit way, the PLM system. For instance, the 'hierarchical constraints' affect the management of access rights to objects and requests; and the 'functional constraints' are implemented in the PLM workflows.

The collaborative problems come in fact from these constraints that, implicitly or indirectly, affect the collaborative activity. For instance, the ones related to informal level of collaboration.

Our aim is not to act on those constraints, but to analyze their consequences by the bias of indicators and to lead to propose actions minimizing their scope and impacts. In this context, the following section presents the collaborative problems identified through a survey.

## 2.2. *Highlights of the collaboration brakes*

### Researches context

The current research is particularly made within the ANCAR-PLM project. This project mainly consists in four working directions:
1. Study of the global integration to IS and the knowledge management;
2. Study of the interoperability and the configuration management solutions;
3. Evaluation, adaptation and integration of the collaboration;
4. PLM deployment methodologies;

The first two working directions are intended to propose a state of the art concerning integration and interoperability dimensions in order to better delimit the necessary conditions for a collaboration development in PLM systems. The third working direction deals with the characterization of the process application within the project partner enterprises. The last working direction is a study on the definition of specific indicators, which will help to establish a control of the collaborative activities in order to anticipate any disruptions in the system. In order to characterize the problems in the implementation and the use of a PLM system, we prepared a survey which was submitted to companies that have already deployed such systems. This study is still ongoing, but the received answers led us to identify some important points.

### Analysis of ANCAR-PLM survey

The analysis of the present answers to the survey highlights some brakes related to the collaborative processes within the PLM system. We globally distinguish two categories of problems:

1. Lack of agility in the processes:

By wanting to control all business processes, the lead team designs processes mobilizing a large volumes of managed product data. This particularly has the following consequences:
- increasing work load by multiplying the validation tasks,
- decreasing reactive capabilities of the enterprise regarding to the customer needs,
- producing side effects on data which are modified by processes
- elevating the maintenance of these systems where processes are managed by lifecycles.

2. Lack of agility in organizational structures:

The management of access rights and requests related to organizational levels is too restrictive and does not allow a global control for user access. This is particularly true in the case of collaboration between different industrial partners where the trust relationship is very important. This problem takes place also place within SMEs where various responsibilities are combined: SME does not ensure project management in a routine way and leads to combine various responsibilities. Indeed, there are not enough actors to fulfil each design role, so most of the actors have various design roles in a project.

Finally, we see that the obstacles to a real collaborative work are mostly related to the level of control. Too much or not enough control (on the process and on the organization) hinders the collaboration. Therefore, in order to facilitate the collaboration there is a need to emphasize by specific indicators the problems encountered in order to allow a balanced and continuous adaptation of the control level. The next section presents a state of the art of collaborative models proposed in the literature.

## 3. Collaboration modelling

The collaboration within PLM systems includes the three levels of collaboration (*Informal, Process/Project and Extended Collaboration*). Both Process/Project and Extended collaboration levels lean particularly on process and organizational meta-models.

This section aims to study the different models addressing collaborative processes in literature.

### 3.1. Collaborative processes

Processes are aimed to model the information dynamic and to define in a coherent way the behaviours of the different handled objects [3]. We talk about collaborative process when one of the process activities is realized by one or more actors. It is necessarily realized by at least two actors, and can gather internal actors to the project and/or to the enterprise, as well as external actors [4]. The needs in coordination of these activities, and afterward in modelling, are very important.

#### Typology of the collaborative processes

We distinguish two main categories of process: repetitive and unique [5]. A repetitive process is executed several times and its modelling presents a normative character: the activities are supposed to be performed in accordance with their description in terms of tasks and their scheduling. A unique process (example: the project) is executed only one time. In this case, one can ask for reasons to model it. It can be for planning reasons (identification of the activities and resources distribution between a group of actors), or for representing a generic structure that can be used as a support to a collaboration tool around a process.

Moreover, we distinguish three process structuring approaches [6]. Certain processes can be defined by a structure that returns completely accounts on activities order, while for others, it is difficult or not very effective to impose all the links between the activities.

The first approach sometimes is termed « mechanistic »: the process role is to define precisely the order and the content of the activities to be performed, and to increase the efficiency and work effectiveness.
In the second approach, termed « systemic », one considers that the activities are components reacting to events. The links between activities are performed by the results: the result of an activity represents a trigger event for other activity. The real sequence of a process instance will correspond to the one of the foreseen ways.
In the third approach, termed « emerging » or of « social constructs », we do not wish to establish any way between activities: It is only a posteriori that eventually we can redraw the activity sequence. Each activity is matched to events able to release it, to interrupt it or to modify its path. The event origin can be external, temporal, or the result of the request of another actor. In this process representation, the sequence is not determined a priori; for example a unique process, in which actors possess latitude in their manner to accomplish an activity.

In addition, we can dissociate the processes varying of those permanent in the time [7]:
-      Permanent processes in the time: the invariants of business processes, i.e. the set of activities that are sufficiently stable and autonomous whatever the business evolutions. These invariants correspond in fact to the big processes of the enterprise;
-      Varying processes: they are easily replaced in the time with changes.

We can also dissociate the processes according to their purpose. Thus, we distinguish two types of processes:
-      Administrative processes: the management processes of information and necessary documents for the accomplishment of different tasks. They are simple and stabilized in the time;
-      Production processes: the processes linked to the services proposed by the organization, assuring its effectiveness. They are generally more complex than the administrative ones (conception, studies, manufactures …)

**Modelling of the collaborative processes**

Several meta-models including the basic process concepts were proposed. Thus, some authors consider that all processes can be modelled in a minimal manner by what one calls sometimes "check flow" (the activities, the activities trigger factors, the activity effects) and the used information [8]. To let the business vision appear, some authors add the objectives, the internal or external actors, certain conditions and resources [9].

In the line of the *norms and standards* dedicated to business processes modelling, we quote the WfMC[3] which proposes a reference model centred on the representation of an automated process, around activity, transition, participant and application concepts [10]. Some authors judge that this model is not rather rigorous for process modelling and implementation [11][12]. Furthermore, currently few products are in accordance with this norm [13]. The norm ISO/IEC15504 [14] aiming to evaluate the processes, proposes an oriented meta-model towards the qualification, with the process concepts, objective, result, category of process and attributes. The norm ENV12204 [14] defines a first version of a unified modelling language. Its basic concepts aim to represent the activity sequence according to different methods (sequence, conditional switching, parallelism, curls). The norm distinguishes between well structured processes, in which the result is determined, semi-structured processes, for which the sequence will be known only at the execution, and non structured processes, for which the result and the sequence are completely unknown [15].

Certain authors also distinguish between the well structured processes where the accent is put on the sequence decomposition (step, tasks…) and the non structured process where only the roles and the resources appear [5], [6].

---

[3] Workflow Management Coalition

Collaborative process modelling is currently in the centre of many researchers preoccupations. It constitutes the principal issue of the modelling efforts.

D. Georgakopoulos [16] treats the process models flexibility problem. He proposes a model allowing reconciling the well structured activities and known in advance and the emerging activities where the process is described progressively. Nevertheless, the proposed model does not furnish any information on the nature and process sequence that allow adjusting the processes.

W. Van Der Aalst and M. Weske [17] propose a global model divided into several parts, each one represents the partner process. This model is compatible with the WfMC norm. It represents a good level of expressiveness and graphic notation. The usage of the Petri networks assures the process coherence.

Wynen [13] proposes a model based on synchronization point concept. It supports data exchange and sharing, and the coordination of the different activities. His model assures process flexibility (possibility to add, eliminate or modify the check flow, the transactional flow, and the message flow). It also assures the dynamic coordination. The used algebraic semantic assures the cooperation activity coherence and execution. Some defects nevertheless can be associated to this model.

Berthier proposes a meta-model called "Actor-Agent-Activity"[5]. It allows activity reuse by including dynamic components, collaborative activities integration, and actor's autonomy. The model also presents a support to emerging processes. It introduces new concepts such as agent, executer, service, Typical Conversation and Social Convention. These two last ones allow representing collaborative activities. Nevertheless the author does not model any of the collaborative aspects, which are at the centre of the processes modelling preoccupations. The meta-model of Berthier brings several openings for constructing process.

The next section proposes an improvement of meta-models in respect of a proposed approach.

## 4. Meta-models improvement

To solve the problems of collaboration identified and described in the previous sections, we propose an approach for improvement based on the meta-models used in the PLM system. This approach can be summarized in two main actions:

1. Proposing an extension of the organizational meta-model in order to be able to monitor and to anticipate collaboration problems.

2. Proposing an extension of the process meta-model in order to be able to unlock collaboration problems.

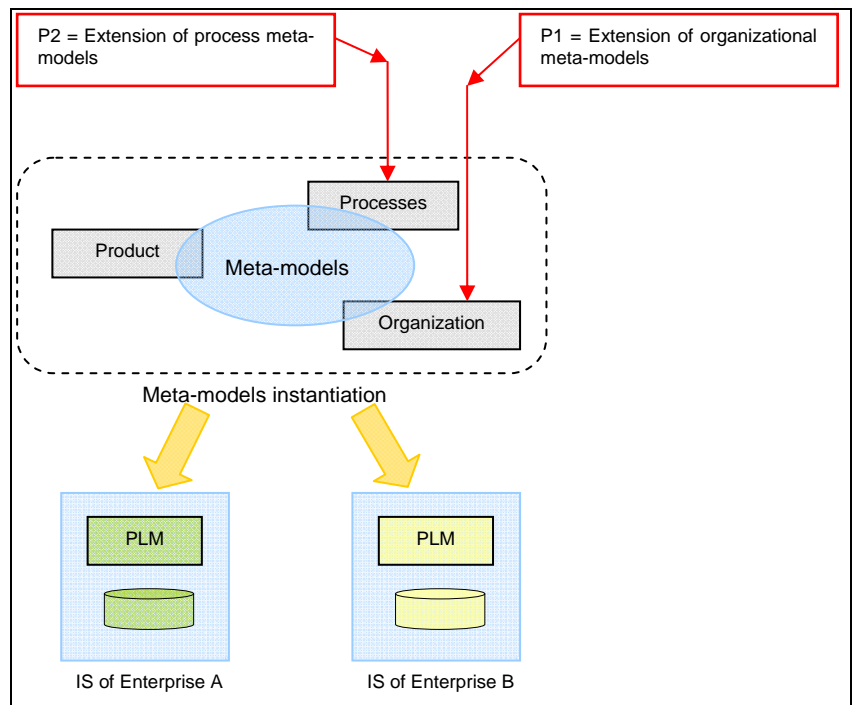

**Figure 1: Meta models improvements within PLM system**

## 4.1. *Deployment of an approach for improvement*

In order to perform the above-mentioned actions, we propose to deploy a **process of observation** of the collaborative activity. The observation aims to the regulation of the activity. It allows obtaining useful information for the adapted sequence of collaborative activity continuation.

In fact, the activity quality can be evaluated in the same manner than software processes, for example with the CMM model [18]. The idea is to modify and to enrich the scenario in which, for instance, certain activities systematically are added or eliminated by the users. To this effect, a track analysis is a posteriori necessity to identify redundant, missing and useless steps. Two sources of observations can be used: the server and the client. Each of these sources necessitates a special instrumentation. A first step incites first of all to orchestrate the PLM system (by the bias of probes) in order to observe abstract elements; and therefore to proceed to their analyze by the bias of monitoring indicators.

The observation process will allow us to identify the different collaborative data. The analysis of these ones will be based on the construction of monitoring indicators.

The methodological approach is proposed in five key steps outlined in the following figure (Figure 2)

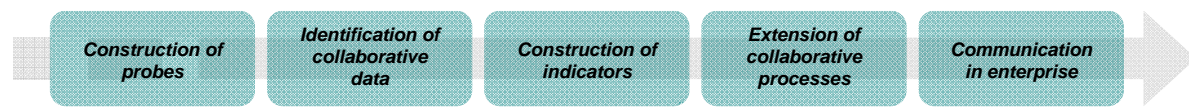

**Figure 2: Key stages of a dynamic improvement of the activity**

These steps summarize how to make the improvement of the cooperation within a PLM system. The following section describes each stage.

**Step 1: Construction of probes**

*S. El Kadiri, P. Pernelle, M. Delattre, A. Bouras*

This stage is a key stage because it allows observing the collaborative activity. The traceability proposed by the PLM system allows arrangement of many traces left by the actor's activity within the system. The observation is realized by probes, which are programs capturing the various elements on the PLM server.

The work in the fields of ILE [4] and especially the approach by the agents [19] [20] identifies several types of observation agents: collector agent, historic agent, structuring agent, statistical agent, visualization agent.
In the context of PLM systems, we propose to use three types of agents:
- Collector agent is responsible for collecting traces from user actions.
- Structuring agent is specialized for grouping, reformatting or annotating the collected traces, to make them more suitable for the construction of indicators.
- Statistical agent is responsible for establishing statistics on the objects' usage.

All these agents, implemented in the PLM system will enable the realization of two actions: identify the collaborative data and build collaborative indicators.

### Step 2: Identification of collaborative data in PLM

Actual PLM systems integrate a set of meta-models (product data, process and organization). Each class instanced is involved in the product development. However, all these elements do not contribute in the same way to the collaborative process. From the meta-data and processes introduced, we propose to identify present objects (class document, specific CAD model…) and flow control in this collaborative process. Hence, we have identified a weight function that is directly driven by observation agents.

$$\delta_c = \sum_{j=1}^{N} \alpha_j A_j \quad (1)$$

$A_i$ is a criterion characterizing a parameter assessable by the statistics agent and $\alpha_j$ a weighting factor.

Some examples of criteria:
- A1 = number of modifications
- A2 = number of access by more than two actors
- A3 = number of tasks where the object is at the output of flow

The construction of a weight function is not an indicator in the strict sense of term but it can be used to restrict the field of survey data to those which are relevant from a collaborative point of view.

### Step 3: Construction of indicators

We propose to construct monitoring indicators in order to analyze and identify gaps fluidity. Their implementation provides a performance management and monitoring of collaborative processes. Indeed, the identification of problems relies on the interpretation of monitoring indicators supplied by the agents of the observation. In this respect, it's advisable to specify the relevant dimensions linking collaborative activity to the manipulated object. This latter corresponds to the set of artefacts used in the collaboration (product information, application, service …). The following list proposes some indicators:
- Number of change requests on a same object;
- Number of validation requests on a same object;
- Time spent on a given task;
- Number of modifications performed on a process;
- Number of non respected times to realize a given task;

---

[4] Interactive Learning Environment

- Time spent on information search on a same object (reactive capacity of the actors);
- The exchange type adopted.

### Step 4: Extension of collaborative processes

The collaborative processes are marked by their instability and by the incompleteness and incoherence of rules which govern them. The need is therefore felt to resort to a virtually permanent process of drafting new games rules to their adaptation and optimization. Indeed, investment in a process of regulatory interactions among the various actors helps to maintain the balance of cooperation. We describe the process of regulation (Figure 3) by:
- Detection, based on the implemented monitoring indicators, of problems or sticking points on the workflow;
- Adaptation, based on the definition of regulatory mechanisms, to propose a solution to the problem;
- Acceptance, based on the cooperation of all stakeholders;
- Implementation of proposed rules and protocols on the workflow model (the adoption of new rules).

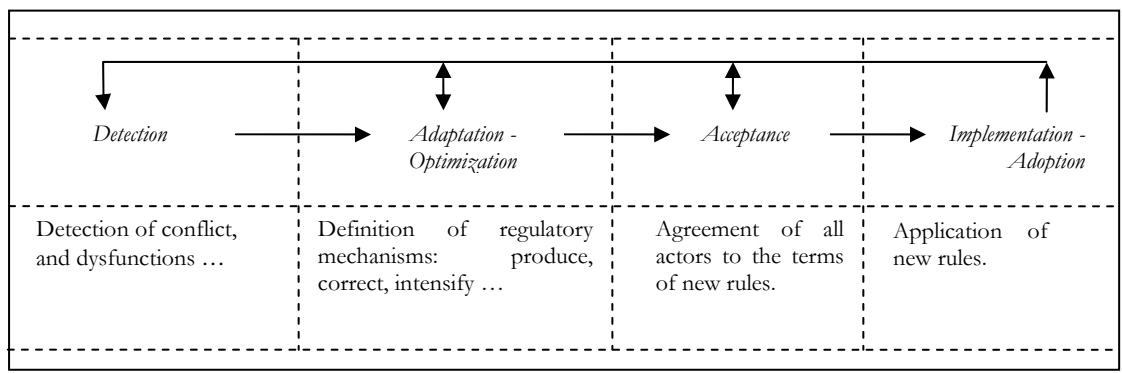

**Figure 3: Progressive model for extension of collaborative processes (Step 4)**

### Step 5: Communication in the enterprise

The different experiences in the implementation of a PLM system show that it is essential to communicate around the developed tools. The collective appropriation of a PLM system reports directly to the quality of the learning process [21]. In fact, the actors must have clear and adequate information on the conduct of collaborative activities to limit their proxies. They must also be able to know the choices adopted and their objective and to perceive the impact on personal and corporate situations.

The myth of the definitive organization and the acquired solutions once for all remain strong. In a context of imbalances and uncertainties, to assure the process agility, organizational structures must remain scalable, unfinished [22] in order to improve the negotiations framework between actors [23]. Communication quality resulted from this methodology, can be evaluated by some performance indicators, including among others the number of system users, user's ability to exploit the resource of the system, degree of use stability. These indicators can allow to assess the maturity of the PLM system, and to lead by the way to improvement initiatives well targeted.

Moreover, the introduction of monitoring indicators and regulatory mechanisms in the proposed methodology, leads us to extend the processes and organizational meta-models in order to support a cooperation approach. So we define the following concepts:

- Indicator: Indicator monitoring with the objective of controlling activities manipulating objects in the system. It is formalized by *an objective*, *a method of calculation*, and *a threshold*.
- Regulator: Its role is to implement the regulation mechanisms and moderation in a collaborative approach. It may be a person or a computer program.
- Rule: It is associated to indicators and defined by the regulator in order to supervise activities.
- I-Orga/Proc/Prod: It is the interface with the organizational/process/product meta models. It corresponds to navigation in each meta model, according to disposed access rights and views.

The extended meta-model is illustrated in Figure 4.

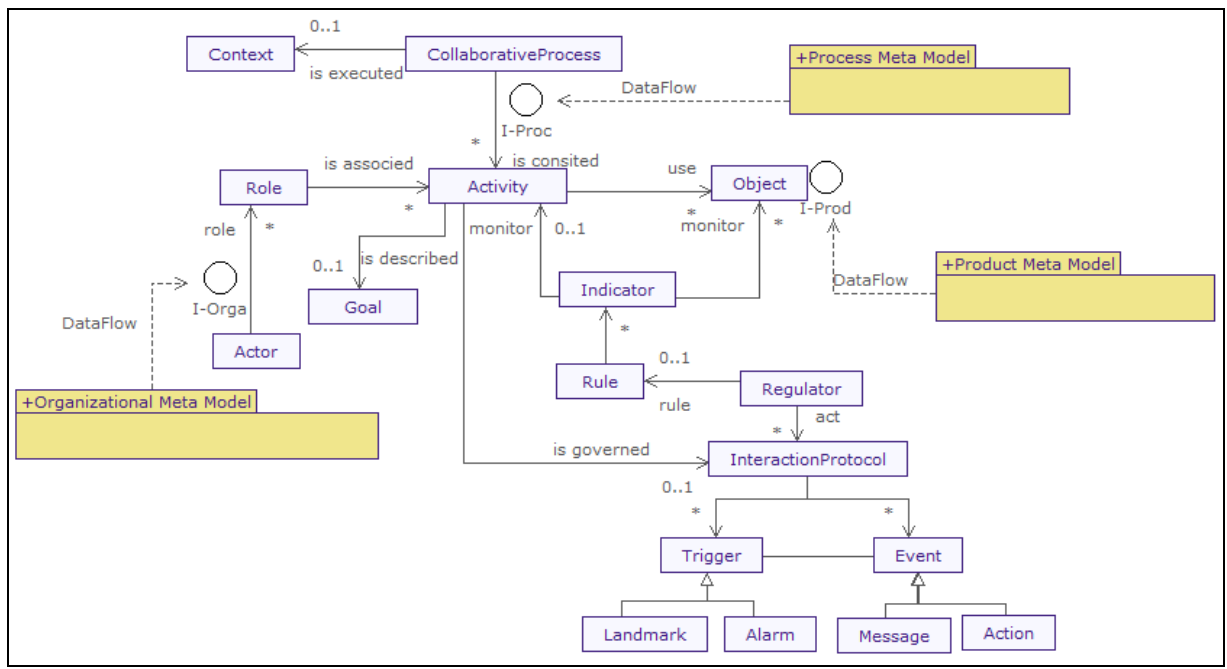

**Figure 4: Extended model of collaborative processes**

## 4.2. Application

The introduction of this method to improve the collaborative process is currently being implemented in an enterprise deploying a PLM system. In this SME, which is a subcontractor company in the plastics sector, the activity is controlled by the responses to request for proposal.

This example of collaborative process involves three sub-processes with several internal and external actors. They are described as follows:

- sub-process 1 : design of the technical characteristics of the product to be developed.
- sub-process 2 : management of internal requests for proposal.
- sub-process 3 : Treatment responses and construction of the final response.

The monitoring of these processes will be based on the indicators proposed in our approach. Some processes are not followed by our indicators to compare the effectiveness of such a method.

# 5.  CONCLUSION

This paper discussed the problems associated with the SMEs collaboration, mainly based on PLM systems. A field study of PLM practices is under way to analyze and identify gaps or fluidity sticking points at the collaborative processes. The exploitation of the first results of this study allowed us to identify two categories of problems: a lack of flexibility in the processes, and a lack of flexible organizational structures.

This led us to propose a methodology for improving collaboration (internal and external) that consists in two decisive actions: the first one explains the construction of monitoring indicators; the second one proposes an extension of model processes to mitigate the previously identified sticking points.

In addition, apart from the implementation of indicators to measure performance, the evolving nature of the proposed methodological approach is part of a process of quality and continuous improvement, enabling the optimization of business processes.